\documentclass[10pt,letterpaper]{article}
\usepackage[top=0.85in,left=2.75in,footskip=0.75in]{geometry}

\usepackage{amsmath,amssymb}

\usepackage{changepage}

\usepackage[utf8x]{inputenc}

\usepackage{textcomp,marvosym}

\usepackage{cite}

\usepackage{nameref,hyperref}

\usepackage[right]{lineno}

\usepackage{microtype}
\DisableLigatures[f]{encoding = *, family = * }

\usepackage[table]{xcolor}

\usepackage{array}

\newcolumntype{+}{!{\vrule width 2pt}}

\newlength\savedwidth

\raggedright
\usepackage[aboveskip=1pt,labelfont=bf,labelsep=period,justification=raggedright,singlelinecheck=off]{caption}

\makeatletter
\renewcommand{\@biblabel}[1]{\quad#1.}
\makeatother

\usepackage{lastpage,fancyhdr,graphicx}
\usepackage{epstopdf}
\fancyheadoffset[L]{2.25in}
\fancyfootoffset[L]{2.25in}
\usepackage{todonotes}
\usepackage{physics}
\usepackage{listings}
\usepackage{jlcode} 

\newcommand{\XX}{\mathcal{X}}
\newcommand{\YY}{\mathcal{Y}}
\newcommand{\ZZ}{\mathcal{Z}}
\newcommand{\Lrm}{\mathrm{L}}
\newcommand{\C}{\mathbb{C}}

\definecolor{jlbase}{HTML}{000000}            
\definecolor{jlkeyword}{HTML}{000000}         
\definecolor{jlliteral}{HTML}{000000}         
\definecolor{jlbuiltin}{HTML}{000000}         
\definecolor{jlcomment}{HTML}{000000}         
\definecolor{jlstring}{HTML}{000000}          
\definecolor{jlbackground}{HTML}{EEEEEE}      
\definecolor{jlrule}{HTML}{000000}            

\makeatletter
\lstdefinestyle{mystyle}{
  basicstyle=%
    \ttfamily
    \lst@ifdisplaystyle\scriptsize\fi
}
\makeatother

\begin{document}
\vspace*{0.2in}

\begin{flushleft}
{\Large
\textbf\newline{QuantumInformation.jl---a Julia package for numerical computation in quantum
information theory}
}
\newline
\\
Piotr Gawron\textsuperscript{$\dagger$},
Dariusz Kurzyk\textsuperscript{$\dagger$},
\L{}ukasz Pawela\textsuperscript{*$\dagger$}
\\
\bigskip
\textbf{$\dagger$}
Institute of Theoretical and Applied Informatics, Polish Academy of Sciences,
Bałtycka 5, 44-100 Gliwice, Poland
\bigskip

$\dagger$ Authors contributed equally to this work.
*lpawela@iitis.pl
\end{flushleft}

\section*{Abstract} Numerical investigations are an important research tool in
quantum information theory. There already exists a~wide range of computational
tools for quantum information theory implemented in various programming
languages. However, there is little effort in implementing this kind of tools in
the \texttt{Julia} language. \texttt{Julia} is a~modern programming language
designed for numerical computation with excellent support for vector and matrix
algebra, extended type system that allows for implementation of elegant
application interfaces and support for parallel and distributed computing.
\texttt{QuantumInformation.jl} is a new quantum information theory library
implemented in \texttt{Julia} that provides functions for creating and analyzing
quantum states, and for creating quantum operations in various representations.
An additional feature of the library is a collection of functions for sampling
random quantum states and operations such as unitary operations and generic
quantum channels.

\section*{Introduction}
Numerical investigations are prevalent in quantum information theory. Numerical
experiments can be used to find counter examples for theorems, to test
hypotheses or to gain insight about quantum objects and operations.

The variety of software that supports investigations in quantum information
theory is very large. Yet there are niches that are not well covered. 
The purpose of \texttt{QuantumInformation.jl} library is to provide
functions to create quantum states, manipulate them with quantum channels,
calculate functionals on these objects and sample
them randomly from various distributions. \texttt{QuantumInformation.jl} package is available
on-line at \url{https://github.com/ZKSI/QuantumInformation.jl} and stored at
Zenodo repository \cite{qizenodo}. It is published under GNU General Public
License v3.0.

\subsection*{Related work} A~comprehensive collection of software related to
quantum mechanics, computation and information can be found at Quantiki
\cite{quantikiqclist}---an on-line resource for quantum information research
community. There exist several notable libraries aimed at numerical and symbolic
computation for quantum information theory. Two \texttt{Mathemetica}
libraries---\texttt{QI} \cite{miszczak2011singular} and \texttt{TRQS}
\cite{miszczak2012generating}---were an inspiration for creation
of~\texttt{QuantumInformation.jl}. Additionally the \texttt{QUANTUM}
\cite{munoz2016quantum} library was implemented in \texttt{Mathematica}.
A~library called \texttt{FEYNMAN} implemented in \texttt{Maple}, described in a
series of papers \cite{radtke2005simulation, radtke2006simulation,
radtke2007simulation, radtke2008simulation, radtke2010simulation}, provides
a~wide variety of functions. The above-mentioned libraries rely on non-free
software and therefore their use can be very limited as use of this software
requires acquiring expensive licenses and its source code cannot be studied by
researchers. Therefore any results obtained using this software rely on trust to 
the companies that produced it. Hence non-free software creates barriers for reproducibility of
scientific results \cite{ince2012case}.

A~widely celebrated and used framework \texttt{QuTiP} \cite{johansson2012qutip,johansson2013qutip}
was written in~\texttt{Python}. \texttt{Python} posses many scientific
computation libraries. It is free software and is widely used for scientific
computation. Nevertheless, as a~general purpose programming language it has its
limits. In~\texttt{Python}, implementations of multidimensional arrays and 
linear
algebra routines are provided by \texttt{NumPy} \cite{numpy} and \texttt{SciPy}
\cite{scipy} respectively. Unfortunately, due to low efficiency of
\texttt{Python}, many of the underling functions are implemented in \texttt{C}
or \texttt{Fortran} programming languages. Therefore, study and development of
these routines is difficult and requires familiarity with these low-level
languages.

\texttt{Julia}\cite{julia}, being a~high-level just-in-time compiled language,
is very efficient and therefore extremely useful for scientific computing. There
are several libraries related to quantum mechanics and quantum information
written in Julia. Those are: \texttt{QuantumInfo.jl} \cite{quantuminfojl},
\texttt{Quantum.jl} \cite{quantumjl} and a~collection of packages developed as
a~part of \texttt{JuliaQuantum} project \cite{juliaquantum}. Unfortunately these
development efforts stalled a couple of years ago. \texttt{JuliaQuantum} project
is very ambitious, but its scope seems to be too large to be implemented fully
in a relatively short amount of time. The only package whose development was
successful is \texttt{QuantumOptics.jl}---a~\texttt{Julia} framework for
simulating open quantum systems \cite{quantopticsjl}. Yet the applicability
scope of this package is different than the one of
\texttt{QuantumInformation.jl}.

\subsection*{Design principles} Our goal while designing \texttt{QuantumInformation.jl} library
was to follow the principles presented in the book ``Geometry of Quantum
States''\cite{bg2008}. We work with column vectors representing kets and row
vectors representing bras. We fix our basis to the computational one. Density
matrices and quantum channels are represented as two-dimensional arrays in the
same fixed basis. This approach allows us to obtain a low level of complexity of
our code, high flexibility and excellent computational efficiency. The design
choices were highly motivated by the properties of the language in which our
library was implemented, namely \texttt{Julia}\cite{bezanson2017julia}.

\texttt{Julia} is a novel scientific programming language mainly influenced by
\texttt{Python}, \texttt{Matlab} and \texttt{Lisp} programming languages. One of
the main concepts widely used in \texttt{Julia} is multiple dispatch
\textit{i.e.} an ability to dispatch function calls to different methods
depending on the types of all function arguments. The multiple dispatch
mechanism together with a~simple yet flexible type system allows to build clean
and easy to use programming interfaces. \texttt{Julia} is just-in-time compiled
to machine code using \texttt{LLVM} \cite{lattner2004llvm} therefore, despite
being a high-level programming language, it can reach computation efficiency
similar to \texttt{C} or \texttt{Fortran}. \texttt{Julia} natively supports
parallel and distributed computing techniques. Therefore it is easy to write
programs for Monte-Carlo sampling in \texttt{Julia}.

In \texttt{Julia} arrays are first class objects \cite{bezanson2014array}, and
linear algebra operations are integrated into the language standard library. The
array system in \texttt{Julia} is designed in a way that minimizes the amount of
memory copying operations during transformations of arrays. \texttt{Julia}
supports various representations of vectors and matrices. For these reasons
a~design decision was made not to create library specific types but to rely on
built-in standard library abstract array types.

The \texttt{QuantumInformation.jl} library was initially developed in \texttt{Julia} 0.6 but
then subsequently it was ported to \texttt{Julia} version 1.0. Part of the
functionality of the library, namely the function that calculates the diamond
norm of a quantum channel relies on \texttt{Convex.jl} library \cite{convexjl}.
Partial traces are implemented using \texttt{TensorOperations.jl} library \cite{tensoroperations} 
that provides basic tensor contractions primitives.

\subsection*{Testing} The \texttt{QuantumInformation.jl} library was tested
using standard \texttt{Julia} framework. Tests where performed using three
distinct approaches. In case of most of the functions the basic properties, such
as e.g. dimensions, norms, hermititicty, positivity, trace are tested, where it was
appropriate. Additionally some test cases where manually computed and used to
verify the obtained results. In the case of methods generating random objects
such as random matrices statistical properties of results are tested. For
example in case of random unitary matrices sampling we test phases distribution
\cite{mezzadri2006generate} of obtained matrices in order to ensure that the
unitary matrices are drawn according to the Haar measure.

\subsection*{Organization of the paper} In the section \textbf{Linear algebra in
Julia}, we describe briefly how the linear algebra routines are implemented in \texttt{Julia}.
Next, in the section \textbf{States and channels}, we
introduce the notions of \emph{quantum states} and \emph{quantum channels}
and we discuss how we implement these concepts in \texttt{Julia}.
Subsequently, the section \textbf{Functionals} focuses on functionals related
with quantum information processing, \textit{i.e.} \emph{trace norm, diamond
norm, entropy, fidelity} or the \emph{PPT criterion}. Afterward, we show the
usage of \texttt{QuantumInformation.jl} for modeling and application of the \emph{quantum
measurements}. The section \textbf{Random quantum objects} introduces
probabilistic measures on quatum states and channels and their implementation in
\texttt{Julia}. Additionally, we introduce some common random matrix ensembles.
In section \textbf{Benchmarks} we provide a comparison, in terms of code clarity and execution speed,
of our library with the latest version of QuTiP \cite{johansson2012qutip,johansson2013qutip}.
Finally, in the section \textbf{Conclusions and future work} we present the final remarks
and outline possible future work.

\section*{Linear algebra in Julia}
A basic construction of vector in \texttt{Julia} creates a~full one-index array
containing elements of a~number type as presented below.
\begin{lstlisting}
julia> x = [0.0, 1.0im]
2-element Array{Complex{Float64},1}:
 0.0+0.0im
 0.0+1.0im
\end{lstlisting}
A transposition of a~column vector returns an object of type
\lstinline|LinearAlgebra.Transpose| as shown below
\begin{lstlisting}
julia> xt = transpose(x)
1×2 LinearAlgebra.Transpose{Complex{Float64},Array{Complex{Float64},1}}:
 0.0+0.0im  0.0+1.0im
\end{lstlisting}
While a~Hermitian conjugate of the same vector returns a \lstinline|LinearAlgebra.Adjoint|
parametrized by the type \lstinline|Array|:
\begin{lstlisting}
julia> xc = [0.0, 1.0im]'
1×2 LinearAlgebra.Adjoint{Complex{Float64},Array{Complex{Float64},1}}:
 0.0-0.0im  0.0-1.0im
\end{lstlisting}
Values of variables \lstinline|xt| and \lstinline|xc| are views of the value of
variable \lstinline|x|. The column and row vectors behave like bras and kets,
for example \lstinline|xc*x| denotes the inner product of `bra' \lstinline|xc|
and `ket' \lstinline|x|, while \lstinline|x*xc| denotes their outer product
resulting in a two-index array.

The linear algebra library in \texttt{Julia} provides standard operations on
matrices and vectors that are designed to take into account the types of the
objects.

\section*{States and channels}\label{sec:states}
In this and the following sections we will denote complex Euclidean spaces
$\C^d$ with $\XX$, $\YY$, $\ZZ$ etc. When needed the dimension of a space $\XX$
will be denoted $\mathrm{dim}(\XX)$. The set of matrices transforming vectors
from $\XX$ to $\YY$ will be denoted $\Lrm(\XX, \YY)$. For simplicity we will
write $\Lrm(\XX) \equiv \Lrm(\XX, \XX)$.

\subsection*{States} By $|\psi\rangle\in\mathcal{X}$ we denote a normed column
vector. Notice that any $|\psi\rangle$ can be expressed as
$|\psi\rangle=\sum_{i=1}^{n} \alpha_i |i\rangle$, where $\sum_{i=1}^{n}
|\alpha_i|^2=1$ and the set $\{|i\rangle\}_{i=1}^{n}$ is the computational
basis.

\begin{lstlisting}
julia> ket(1,2)
2-element Array{Complex{Float64},1}:
 1.0 + 0.0im
 0.0 + 0.0im

julia> (1/sqrt(2)) * (ket(1,2) + ket(2,2))
2-element Array{Complex{Float64},1}:
 0.7071067811865475 + 0.0im
 0.7071067811865475 + 0.0im
\end{lstlisting}
According to common academic convention, we count the indices of states starting
from~one. Following the standard Dirac notation the symbol $\bra{\psi}$ denotes
the row vector dual to $\ket{\psi}$. Therefore $\ket{\psi}=\bra{\psi}^\dagger$,
where the symbol ${}^\dagger$ denotes the Hermitian conjugation.
\begin{lstlisting}
julia> bra(2,3)
1×3 LinearAlgebra.Adjoint{Complex{Float64},Array{Complex{Float64},1}}:
 0.0-0.0im  1.0-0.0im  0.0-0.0im
\end{lstlisting}

The inner product of $\ket{\phi}, \ket{\psi} \in \XX$ is denoted by
$\langle\psi|\phi\rangle$ and the norm is defined as
$\|\ket{\phi}\|=\sqrt{\langle\phi|\phi\rangle}$.
\begin{lstlisting}
julia> ψ=(1/sqrt(2)) * (ket(1,2) + ket(2,2))
2-element Array{Complex{Float64},1}:
 0.7071067811865475 + 0.0im
 0.7071067811865475 + 0.0im

julia>  ϕ=(1/2) * ket(1,2) + (sqrt(3)/2) * ket(2,2)
2-element Array{Complex{Float64},1}:
 0.5 + 0.0im
 0.8660254037844386 + 0.0im

julia> ϕ' * ψ
 0.9659258262890682 + 0.0im

julia> sqrt(ϕ' * ϕ)
 0.9999999999999999 + 0.0im
\end{lstlisting}
The form $|{\psi}\rangle\langle{\phi}| \in \Lrm(\XX, \YY)$ denotes outer product
of $|{\psi}\rangle \in \YY$ and $|{\phi}\rangle \in \XX$.
\begin{lstlisting}
julia> ketbra(2,3,4)
4×4 Array{Complex{Float64},2}:
 0.0+0.0im  0.0+0.0im  0.0+0.0im  0.0+0.0im
 0.0+0.0im  0.0+0.0im  1.0+0.0im  0.0+0.0im
 0.0+0.0im  0.0+0.0im  0.0+0.0im  0.0+0.0im
 0.0+0.0im  0.0+0.0im  0.0+0.0im  0.0+0.0im
\end{lstlisting}
Specifically, $|{\psi}\rangle\langle{\psi}|$ is a rank-one projection operator
called a \emph{pure state}. Generally, any \emph{quantum state} $\rho$ can be
expressed as $\rho=\sum_{i=1}^{n} q_i |\psi_i\rangle\langle \psi_i|$, where 
$\sum_{i=1}^{n}
q_i=1$ and $|\psi_i\rangle\langle \psi_i|$ are rank-one projectors. 
Notice that $\rho$ is a trace-one positive semi-definite linear operator
\emph{i.e.}: $\rho=\rho^\dagger$, $\rho\geq 0$ and $\mathrm{tr}{\rho}=1$.
\begin{lstlisting}
julia> proj(ψ)
2×2 Array{Complex{Float64},2}:
 0.5+0.0im  0.5+0.0im
 0.5+0.0im  0.5+0.0im
\end{lstlisting}

For convenience, the \texttt{QuantumInformation.jl} library provides the 
implementations of maximally 
mixed, maximally entangled and Werner states.
\begin{lstlisting}
julia> max_entangled(4)
4-element reshape(::Diagonal{Complex{Float64},Array{Complex{Float64},1}},4) 
with eltype Complex{Float64}:
 0.7071067811865475 + 0.0im
 0.0 + 0.0im
 0.0 + 0.0im
 0.7071067811865475 + 0.0im

julia> max_mixed(4)
4×4 Array{Float64,2}:
0.25  0.0   0.0   0.0
0.0   0.25  0.0   0.0
0.0   0.0   0.25  0.0
0.0   0.0   0.0   0.25

julia> werner_state(4, 0.4)
4×4 Array{Complex{Float64},2}:
 0.35+0.0im   0.0+0.0im   0.0+0.0im   0.2+0.0im
 0.0+0.0im  0.15+0.0im   0.0+0.0im   0.0+0.0im
 0.0+0.0im   0.0+0.0im  0.15+0.0im   0.0+0.0im
 0.2+0.0im   0.0+0.0im   0.0+0.0im  0.35+0.0im
\end{lstlisting}

\subsection*{Non-standard matrix transformations} We will now introduce
reshaping operators, which map matrices to vectors and vice versa. We start with
the mapping
$\mathrm{res}:\mathrm{L}(\mathcal{X,Y})\to\mathcal{Y}\otimes\mathcal{X}$, which
transforms the matrix $\rho$ into a vector row by row. More precisely, for
dyadic operators $|\psi\rangle\langle\phi|$, where $|\psi\rangle \in
\mathcal{Y}$, $|\phi\rangle \in \mathcal{X}$ the operation $\mathrm{res}$ is
defined as
$\mathrm{res}(|\psi\rangle\langle\phi|)=|\psi\rangle|\overline{\phi}\rangle$ and
can be uniquely extend to the whole space $\mathrm{L}(\mathcal{X,Y})$ by
linearity.
\begin{lstlisting}
julia> res(ketbra(1,2,2))
4-element reshape(::LinearAlgebra.Transpose{Complex{Float64},Array{Complex{Float64},2}}, 4) 
with eltype Complex{Float64}:
 0.0 + 0.0im
 1.0 + 0.0im
 0.0 + 0.0im
 0.0 + 0.0im
\end{lstlisting}
The inverse operation to $\mathrm{res}$ is
$\mathrm{unres}:\mathcal{Y}\otimes\mathcal{X}\to \mathrm{L}(\mathcal{X,Y}) $,
which transforms the vector into a matrix. It is defined as the unique linear
mapping satisfying $\rho=\mathrm{unres}(\mathrm{res}(\rho))$.
\begin{lstlisting}
julia> unres(res(ketbra(1,2,2)))
2×2 LinearAlgebra.Transpose{Complex{Float64},Base.ReshapedArray{Complex{Float64},2,LinearAlgebra.
Transpose{Complex{Float64},Array{Complex{Float64},2}},
Tuple{Base.MultiplicativeInverses.SignedMultiplicativeInverse{Int64}}}}:
 0.0+0.0im  1.0+0.0im
 0.0+0.0im  0.0+0.0im
\end{lstlisting}
Let us recall that trace is a mapping 
$\mathrm{Tr}:\mathrm{L}(\mathcal{X})\to \mathbb{C},$ given by
$\mathrm{Tr}:\rho\mapsto\sum_{i=1}^{\mathrm{dim}(\mathcal{X})}\langle
e_i|\rho|e_i\rangle$, where $\{|e_i\rangle \}$ is an orthonormal basis of
$\mathcal{X}$. According to this, \emph{partial trace} is a mapping
$\mathrm{Tr}_{\mathcal{X}}: \mathrm{L}(\mathcal{X}\otimes\mathcal{Y}) \to
\mathrm{L}(\mathcal{Y})$ such that $\mathrm{Tr}_{\mathcal{X}}: \rho_A\otimes
\rho_B \mapsto \rho_B \mathrm{Tr}(\rho_A)$, where $\rho_A\in
\mathrm{L}(\mathcal{X})$, $\rho_B\in \mathrm{L}(\mathcal{Y})$. As this is a
linear map, it may be uniquely extended to the case of operators which are not
in a tensor product form.
\begin{lstlisting}
julia> ρ = [0.25 0.25im; -0.25im 0.75]
2×2 Array{Complex{Float64},2}:
 0.25+0.0im   0.0+0.25im
-0.0-0.25im   0.75+0.0im

julia> σ = [0.4 0.1im; -0.1im 0.6]
2×2 Array{Complex{Float64},2}:
 0.4+0.0im   0.0+0.1im
-0.0-0.1im   0.6+0.0im

julia> ptrace(ρ ⊗ σ, [2, 2], [2])
2×2 Array{Complex{Float64},2}:
 0.25+0.0im    0.0+0.25im
 0.0-0.25im    0.75+0.0im
\end{lstlisting}
Matrix transposition is a mapping ${}^T:\mathrm{L}(\mathcal{X,Y}) \to
\mathrm{L}(\mathcal{Y,X})$ such that $\left(\rho^T \right)_{ij} = \rho_{ji}$,
where $\rho_{ij}$ is a $i$-th row, $j$-th column element of matrix $\rho$.
Following this, we may introduce \emph{partial transposition}
${}^{\Gamma_B}:
\mathrm{L}(\mathcal{X}_A \otimes \mathcal{X}_B, \mathcal{Y}_A \otimes \mathcal{Y}_B) \to
\mathrm{L}(\mathcal{X}_A \otimes \mathcal{Y}_B, \mathcal{Y}_A \otimes \mathcal{X}_B)$, 
which for a~product
state $\rho_A\otimes\rho_B$ is given by ${}^{\Gamma_B}:
\rho_A\otimes\rho_B\mapsto\rho_A\otimes\rho_B^T$. The definition of partial
transposition can be uniquely extended for all operators from linearity.
\begin{lstlisting}
julia> ptranspose(ρ ⊗ σ, [2, 2], [1])
4×4 Array{Complex{Float64},2}:
 0.1+0.0im       0.0+0.025im     0.0-0.1im      0.025-0.0im
 0.0-0.025im     0.15+0.0im     -0.025+0.0im    0.0-0.15im
 0.0+0.1im      -0.025+0.0im     0.3+0.0im      0.0+0.075im
 0.025-0.0im     0.0+0.15im      0.0-0.075im    0.45+0.0im
\end{lstlisting}
For given multiindexed matrix $\rho_{(m,\mu),(n,\nu)}=\langle
m \mu|\rho|n \nu\rangle$, the reshuffle operation is defined as
$\rho^R_{(m,\mu),(n,\nu)}=\rho_{(m,n),(\mu,\nu)}$.
\begin{lstlisting}
julia> reshuffle(ρ ⊗ σ)
4×4 Array{Complex{Float64},2}:
 0.1+0.0im    0.0+0.025im     0.0-0.025im     0.15+0.0im
 0.0+0.1im   -0.025+0.0im     0.025-0.0im     0.0+0.15im
 0.0-0.1im    0.025-0.0im    -0.025+0.0im     0.0-0.15im
 0.3+0.0im    0.0+0.075im     0.0-0.075im     0.45+0.0im
\end{lstlisting}

\subsection*{Channels} Physical transformations of quantum states into quantum
states are called quantum channels \textit{i.e.} linear Completely Positive
Trace Preserving (CP-TP) transformations. Probabilistic transformations of
quantum states are called quantum operations and mathematically they are defined
as linear Completely Positive Trace Non-increasing (CP-TNI) maps. For the sake
of simplicity we will refer to both CP-TP and CP-TNI maps as quantum channels
when it will not cause confusion.

There exists various representations of quantum channels such as:
\begin{itemize}
\item Kraus operators,
\item natural representation, also called superoperator representation,
\item Stinespring representation,
\item Choi-Jamio{\l}kowski matrices, sometimes called dynamical matrices.
\end{itemize}

Formally, properties of quantum channels can be stated as
follows~\cite{watrous2018theory}. First, we introduce the notion of
\emph{superoperator} as a linear mapping acting on linear operators
$\mathrm{L}(\mathcal{X})$ and transforming them into operators acting on
$\mathrm{L}(\mathcal{Y})$. The set of all such mapping will be denoted by
$\mathrm{T}(\mathcal{X},\mathcal{Y})$ and
$\mathrm{T}(\mathcal{X})\equiv\mathrm{T}(\mathcal{X},\mathcal{X})$. In
mathematical terms, a quantum channel is a superoperator
$\Phi:\mathrm{L}(\mathcal{X})\to \mathrm{L}(\mathcal{Y})$ that is
\begin{itemize}
	\item \emph{trace-preserving} ($\forall \rho\in \mathrm{L}(\mathcal{X})\quad
	\mathrm{Tr}(\Phi(\rho))=\mathrm{Tr}(\rho)$) and
	\item \emph{completely positive}
	($\forall \mathcal{Z} \forall \rho \in \mathrm{L}(\mathcal{X\otimes Z}),
	\rho\geq 0, \Phi\otimes\mathbb{I}_{\mathrm{L}(\mathcal{Z})}(\rho)\geq 0$).
\end{itemize}

The product of superoperators $\Phi_1\in
\mathrm{T}(\mathcal{X}_1,\mathcal{Y}_1)$, $\Phi_2\in
\mathrm{T}(\mathcal{X}_2,\mathcal{Y}_2)$ is a mapping $\Phi_1\otimes\Phi_2\in
T(\mathcal{X}_1\otimes\mathcal{X}_2,\mathcal{Y}_1\otimes\mathcal{Y}_2)$ that
satisfies
$(\Phi_1\otimes\Phi_2)(\rho_1\otimes\rho_2)=\Phi_1(\rho_1)\otimes\Phi_2(\rho_2)$.
For the operators that are not in a tensor product form this notion can be
uniquely extended from linearity.

According to Kraus' theorem, any completely positive trace-preserving (CP-TP)
map $\Phi$ can always be written as $\Phi(\rho)=\sum_{i=1}^r K_i \rho
K_i^\dagger$ for some set of operators $\{K_i\}_{i=1}^r$ satisfying
$\sum_{i=1}^r K_i^\dagger K_i = \mathbb{I}_\mathcal{X}$, where $r$ is the rank
of superoperator $\Phi$.

Another way to represent the quantum channel is based on Choi-Jamiołkowski
isomorphism. Consider mapping $J:\mathrm{T}(\mathcal{X,Y})\to
\mathrm{L}(\mathcal{Y}\otimes\mathcal{X})$ such that
$J(\Phi)=(\Phi\otimes\mathbb{I}_{\mathrm{L}(\mathcal{X})})
(\mathrm{res}(\mathbb{I}_{\mathcal{X}})
\mathrm{res}(\mathbb{I}_{\mathcal{X}})^\dagger)$.
Equivalently
$J(\Phi)=\sum_{i,j=1}^{\mathrm{dim(\mathcal{X})}}\Phi(|i\rangle\langle
j|)\otimes|i\rangle\langle j|$. The action of a superoperator in the Choi
representation is given by
$\Phi(\rho)=\mathrm{Tr}_\mathcal{X}(J(\Phi)(\mathbb{I}_\mathcal{Y}\otimes\rho^T))$.

The natural representation of a quantum channel $\mathrm{T}(\XX, \YY)$ is a
mapping $\mathrm{res}(\rho) \mapsto \mathrm{res}(\Phi(\rho))$. It is represented
by a matrix $K(\Phi) \in \Lrm(\XX \otimes \XX, \YY \otimes \YY)$ for which the following holds
\begin{equation}
K(\Phi) \mathrm{res}(\rho) = \mathrm{res}(\Phi(\rho)),
\end{equation}
for all $\rho \in \Lrm(\XX)$.

Let $\XX, \YY$ and $\ZZ$ be a complex Euclidean spaces. The action of the
Stinespring representation of a quantum channel $\Phi\in \mathrm{T}(\XX,\YY)$ on
a state $\rho\in \Lrm(\XX)$ is given by
\begin{equation}
\Phi(\rho)=\Tr_\ZZ(A\rho A^\dagger),
\end{equation}
where $A\in\Lrm(\XX,\YY\otimes\ZZ)$.

We now briefly describe the relationships among channel representations
\cite{watrous2018theory}. Let $\Phi\in \mathrm{T}(\mathcal{X}, \mathcal{Y})$ 
be a quantum channel which can be written in the Kraus representation as
\begin{equation}
\Phi(\rho)=\sum_{i=1}^r K_i \rho K_i^\dagger,
\end{equation}  
where $\{K_i\}_{i=1}^r$ are Kraus operators satisfying $\sum_{i=1}^r K_i^\dagger
K_i =
\mathbb{I}_\mathcal{X}$. According to this assumption, $\Phi$ can be represented
in 
\begin{itemize}
	\item Choi representation as 
	\begin{equation}
	J(\Phi)=\sum_{i=1}^r \mathrm{res}(K_i)\mathrm{res}(K_i^\dagger),
	\end{equation}
	\item natural representation as 
	\begin{equation}
	K(\Phi)=\sum_{i=1}^r K_i\otimes K_i^*,
	\end{equation}
	\item Stinespring representation as
	\begin{equation}
	\Phi(\rho)=\tr_\ZZ(A\rho A^\dagger),
	\end{equation}
	where $A=\sum_{i=1}^r K_i\otimes \ket{e_i}$ and $\ZZ=\C^r$.
\end{itemize}


In \texttt{QuantumInformation.jl} states and channels are always represented in the
computational basis therefore channels are stored in the memory as either
vectors of matrices in case of Kraus operators or matrices in other cases. In
\texttt{QuantumInformation.jl} quantum channels are represented by a~set of types deriving from
an abstract type \lstinline|AbstractQuantumOperation{T}| where type parameter
\lstinline|T| should inherit from \lstinline|AbstractMatrix{<:Number}|. Every
type inheriting from \lstinline|AbstractQuantumOperation{T}| should contain
fields \lstinline|idim| and \lstinline|odim| representing the dimension of input
and output space of the quantum channel.

Two special types of channels are implemented: \lstinline|UnitaryChannel| and
\lstinline|IdentityChannel| that can transform ket vectors into ket vectors.

\subsubsection*{Constructors}
Channel objects can be constructed from matrices that represent them, as shown
in the following listing
\begin{lstlisting}
julia> γ=0.4
 0.4

julia> K0 = Matrix([1 0; 0 sqrt(1-γ)])
2×2 Array{Float64,2}:
 1.0  0.0
 0.0  0.774597

julia> K1 = Matrix([0 sqrt(γ); 0 0])
2×2 Array{Float64,2}:
 0.0  0.632456
 0.0  0.0

julia> Φ = KrausOperators([K0,K1])
KrausOperators{Array{Float64,2}}
dimensions: (2, 2)
 [1.0 0.0; 0.0 0.774597]
 [0.0 0.632456; 0.0 0.0]

julia> iscptp(Φ)
 true
\end{lstlisting}
There are no checks whether a matrix represents a valid CP-TP or CP-TNI map,
because this kind of verification is costly and requires potentially expensive
numerical computation. Function such as \lstinline|iscptp()|, and
\lstinline|iscptni()| are provided to test properties of supposed quantum
channel or quantum operation.

\subsubsection*{Conversion} Conversions between all quantum channel types,
\textit{i.e.} these that derive from \lstinline|AbstractQuantumOperation{T}|
are implemented. The users are not limited by any single channel representation
and can transform between representations they find the most efficient or
suitable for their purpose.

\begin{lstlisting}
julia> Ψ1 = convert(SuperOperator{Matrix{ComplexF64}}, Φ)
SuperOperator{Array{Complex{Float64},2}}
dimensions: (2, 2)
Complex{Float64}
 [1.0+0.0im 0.0+0.0im 0.0+0.0im 0.4+0.0im; 
 0.0+0.0im 0.774597+0.0im 0.0+0.0im 0.0+0.0im; 
 0.0+0.0im 0.0+0.0im 0.774597+0.0im 0.0+0.0im; 
 0.0+0.0im 0.0+0.0im 0.0+0.0im 0.6+0.0im]

julia> Ψ2 = convert(DynamicalMatrix{Matrix{Float64}}, Φ)
DynamicalMatrix{Array{Float64,2}}
dimensions: (2, 2)
 [1.0 0.0 0.0 0.774597; 
 0.0 0.4 0.0 0.0; 
 0.0 0.0 0.0 0.0; 
 0.774597 0.00.0 0.6]

julia> Ψ3 = convert(Stinespring{Matrix{Float64}}, Φ)
Stinespring{Array{Float64,2}}
dimensions: (2, 2)
 [0.0 0.0; 
-1.82501e-8 0.0; 
 … ; 
 0.0 0.0; 
 0.0 -0.774597]
\end{lstlisting}

\subsubsection*{Application} Channels can act on pure and mixed states 
represented by vectors and matrices respectively. Channels are callable and therefore mimic
application of a~function on a~quantum state.
\begin{lstlisting}
julia> ρ1=ψ * ψ'
2×2 Array{Complex{Float64},2}:
 0.5+0.0im  0.5+0.0im
 0.5+0.0im  0.5+0.0im

julia> Φ(ρ1)
2×2 Array{Complex{Float64},2}:
 0.7+0.0im        0.387298+0.0im
 0.387298+0.0im   0.3+0.0im

julia> Ψ1(ρ1)
2×2 Array{Complex{Float64},2}:
 0.7+0.0im        0.387298+0.0im
 0.387298+0.0im   0.3+0.0im

julia> Φ(ψ)
2×2 Array{Complex{Float64},2}:
 0.7+0.0im        0.387298+0.0im
 0.387298+0.0im   0.3+0.0im
\end{lstlisting}

\subsubsection*{Composition} Channels can be composed in parallel or in
sequence. Composition in parallel is done using \lstinline|kron()| function or
the overloaded $\otimes$ operator. Composition in sequence can be done in two
ways either by using \texttt{Julia} built-in function composition operator
$(f\circ g)(\cdot)=f(g)(\cdot)$ or by using multiplication of objects inheriting
from \lstinline|AbstractQuantumOperation{T}| abstract type.
\begin{lstlisting}
julia> ρ2=ϕ * ϕ'
2×2 Array{Complex{Float64},2}:
 0.25+0.0im      0.433013+0.0im
 0.433013+0.0im  0.75+0.0im

julia> (Φ ⊗ Φ)(ρ1 ⊗ ρ2)
4×4 Array{Complex{Float64},2}:
 0.385+0.0im     0.234787+0.0im  0.213014+0.0im  0.129904+0.0im
 0.234787+0.0im  0.315+0.0im     0.129904+0.0im  0.174284+0.0im
 0.213014+0.0im  0.129904+0.0im  0.165+0.0im     0.100623+0.0im
 0.129904+0.0im  0.174284+0.0im  0.100623+0.0im  0.135+0.0im

julia> (Ψ1 ∘ Ψ2)(ρ1)
2×2 Transpose{Complex{Float64},Array{Complex{Float64},2}}:
0.82+0.0im   0.3+0.0im
0.3+0.0im  0.18+0.0im
\end{lstlisting}

\section*{Functionals}

\subsubsection*{Trace norm and distance} Let $\rho, \sigma \in
\mathrm{L}(\mathcal{X})$. The \emph{trace norm} is defined as $\|\rho\|_1 =
\mathrm{Tr} \sqrt{\rho\rho^\dagger}$ and the trace distance is defined as
$D_1(\rho,\sigma)=\frac{1}{2}\|\rho-\sigma\|_1$.
\begin{lstlisting}
julia> ψ=(1/sqrt(2)) * (ket(1,2) + ket(2,2))
2-element Array{Complex{Float64},1}:
 0.7071067811865475 + 0.0im
 0.7071067811865475 + 0.0im

ϕ=(1/2) * ket(1,2) + (sqrt(3)/2) * ket(2,2)
2-element Array{Complex{Float64},1}:
 0.5 + 0.0im
 0.8660254037844386 + 0.0im

julia> ρ=proj(ψ)
2×2 Array{Complex{Float64},2}:
 0.5+0.0im  0.5+0.0im
 0.5+0.0im  0.5+0.0im

julia> σ=proj(ϕ)
2×2 Array{Complex{Float64},2}:
 0.25+0.0im       0.433013+0.0im
 0.433013+0.0im   0.75+0.0im

julia> norm_trace(ρ)
 1.0

julia> trace_distance(ρ, σ)
 0.2588190451025207 + 0.0im
\end{lstlisting}
\subsubsection*{Hilbert--Schmidt norm and distance}
The \emph{Hilbert--Schmidt} norm and distance defined by
$\|\rho\|_{HS}=\sqrt{\mathrm{Tr}\rho^\dagger \rho}$ and
$D_{HS}(\rho,\sigma)=\frac{1}{2}\|\rho-\sigma\|_{HS}$, respectively, can be used as follows
\begin{lstlisting}
julia> norm_hs(ρ)
 0.9999999999999998

julia> hs_distance(ρ, σ)
 0.36602540378443854
\end{lstlisting}
\subsubsection*{Fidelity and superfidelity}
\emph{Fidelity} is a measure of distance of quantum states. It is an example of 
a
distance measure which is not a metric on the space of quantum states. The
fidelity of two quantum states $\rho, \sigma \in \mathrm{L}(\mathcal{X})$ is 
given by
$F(\rho,\sigma)=\|\sqrt{\rho}\sqrt{\sigma}\|_1$
\begin{lstlisting}
julia> fidelity_sqrt(ρ, σ)
 0.9659258262890682

julia> fidelity(ρ, σ)
 0.9330127018922192

julia> fidelity(ψ, σ)
 0.9330127018922191

julia> fidelity(ρ, ϕ)
 0.9330127018922191 

julia> fidelity(ψ, ϕ)
 0.9330127018922192
\end{lstlisting}
\emph{Superfidelity} is an upper bound on the fidelity of two quantum states
It is defined by
$G(\rho, \sigma) = \mathrm{Tr}\rho \sigma + \sqrt{1 - \mathrm{Tr}\rho^2}
\sqrt{1-\mathrm{Tr} \sigma^2}$.
\begin{lstlisting}
julia> superfidelity(ρ, σ)
 0.9330127018922193
\end{lstlisting}
\subsubsection*{Diamond norm}
In order to introduce the \emph{diamond norm}, we first introduce the notion of
the
induced trace norm. Given $\Phi \in \mathrm{T}(\mathcal{X}, \mathcal{Y})$ we 
define its
induced trace
norm as $\| \Phi \|_1 = \mathrm{max} \left\{ \| \Phi(X) \|_1: X \in 
L(\mathcal{X}), \| X
\|_1 \leq 1 \right\}$.
The diamond norm of $\Phi$ is defined as
$ \| \Phi \|_\diamond = \| \Phi \otimes \mathbb{I}_{\Lrm(\YY)} \|_1$.
One important property of the diamond norm is that for Hermiticity-preserving
$\Phi \in \mathrm{T}(\mathcal{X}, \mathcal{Y})$ we obtain
$\| \Phi \|_\diamond = \max \left\{ \left\| (\Phi \otimes 
\mathbb{I}_{\Lrm(\YY)})
	\left(|\psi\rangle\langle\psi| \right )\right\|_1: |\psi\rangle \in 
	\mathcal{X} \otimes
	\mathcal{Y},
	\langle\psi|\psi\rangle=1 \right\}$.
\begin{lstlisting}
julia> K0 = Matrix([1 0; 0 sqrt(1-γ)])
2×2 Array{Float64,2}:
 1.0  0.0
 0.0  0.774597

julia> K1 = Matrix([0 sqrt(γ); 0 0])
2×2 Array{Float64,2}:
 0.0  0.632456
 0.0  0.0

julia> Φ = KrausOperators([K0,K1])
KrausOperators{Array{Float64,2}}
dimensions: (2, 2)
 [1.0 0.0; 0.0 0.774597]
 [0.0 0.632456; 0.0 0.0]

julia>  L0 = Matrix([1 0; 0 sqrt(1-γ)])
2×2 Array{Float64,2}:
 1.0  0.0
 0.0  0.774597

julia> L1 = Matrix([0 0; 0 sqrt(γ)])
2×2 Array{Float64,2}:
 0.0  0.0
 0.0  0.632456

julia> Ψ = KrausOperators([K0,K1])
KrausOperators{Array{Float64,2}}
dimensions: (2, 2)
 [1.0 0.0; 0.0 0.774597]
 [0.0 0.632456; 0.0 0.0]

julia> norm_diamond(Φ)
 1.0000000077706912

julia> diamond_distance(Φ, Ψ)
-5.258429449675825e-7
\end{lstlisting}

Diamond norm and diamond distance are implemented using the \texttt{Convex.jl} 
\texttt{Julia} package \cite{convexjl}.

\subsubsection*{Shannon entropy and von Neumann entropy} \emph{Shannon entropy}
is defined for a probability vector $p$ as $H(\mathrm{p})=-\sum_{i=1}^n
p_i\log_2 p_i$. We also provide an implementation for the point Shannon entropy.
It is defined as $h(a) = -a \log a - (1-a)\log(1-a)$.
\begin{lstlisting}
julia> p = [0.3, 0.2, 0.5]
3-element Array{Float64,1}:
 0.3
 0.2
 0.5

julia> shannon_entropy(p)
 1.0296530140645737

julia> shannon_entropy(0.5)
 0.6931471805599453
\end{lstlisting}

For a quantum system described by a state $\rho$, the \emph{von Neumann entropy}
is $S(\rho)=-\mathrm{tr} \rho \log \rho$. Let $\lambda_i$,  $0\leq i < n$ be the
eigenvalues of $\rho$, then $S(\rho)$ can be written as $S(\rho)=-\sum_{i=1}^{n}
\lambda_i \log \lambda_i$.
\begin{lstlisting}
julia> ρ = [0.25 0.25im; -0.25im 0.75]
2×2 Array{Complex{Float64},2}:
 0.25+0.0im   0.0+0.25im
 1-0.0-0.25im  0.75+0.0im

julia> σ = [0.4 0.1im; -0.1im 0.6]
2×2 Array{Complex{Float64},2}:
 0.4+0.0im   0.0+0.1im
-0.0-0.1im  0.6+0.0im

julia> vonneumann_entropy(0.4 * ρ + 0.6 * σ)
 0.5869295208554555
\end{lstlisting}

\subsubsection*{Distinguishability between two quantum states} One of the
measure of distinguishability between two quantum states is the \emph{quantum
relative entropy}, called also Kullback–Leibler divergence, defined as
$S(\rho\|\sigma)=-\mathrm{Tr}\rho\log\sigma + \mathrm{Tr}\rho\log\rho$
\begin{lstlisting}
julia> relative_entropy(ρ, σ)
 0.11273751829075163

julia> kl_divergence(ρ, σ)
 0.11273751829075163
\end{lstlisting}

Another type of measure of distinguishability between two quantum state is
\emph{quantum Jensen–Shannon divergence} given by
$QJS(\rho,\sigma)=S\left(\frac{1}{2}\rho+\frac{1}{2}\sigma\right)-
\left(\frac{1}{2}S(\rho)+\frac{1}{2}S(\sigma)\right)$.
\begin{lstlisting}
julia> js_divergence(ρ, σ)
 0.1252860912303596
\end{lstlisting}

\emph{The Bures distance} defines an infinitesimal distance between quantum
states, and it is defined as $D_B=\sqrt{2(1-\sqrt{F(\rho,\sigma)})}$. The value
related with Bures distance is the Bures angle
$D_A(\rho,\sigma)=\arccos(\sqrt{F(\rho,\sigma)})$
\begin{lstlisting}
julia> bures_distance(ρ, σ)
 0.24867555729886728

julia> bures_angle(ρ, σ)
 0.2493208055929498
\end{lstlisting}

\subsubsection*{Quantum entanglement}
One of the entanglement measures is \emph{negativity} defined as
$\mathrm{N}(\rho)=\frac{\|\rho^{T_A}\|_1-1}{2}$.
\begin{lstlisting}
julia> negativity(ρ ⊗ σ, [2, 2], 2)
-0.0

julia> negativity(proj((1/sqrt(2)*(ket(1,2) ⊗ ket(1,2)-ket(2,2) ⊗ ket(2,2)))), [2, 2], 2)
 0.4999999999999999

julia> log_negativity(ρ ⊗ σ, [2, 2], 2)
-1.1102230246251565e-16
\end{lstlisting}

\emph{Positive partial transpose} (the Peres–Horodecki criterion) is a necessary
condition of separability of the joint state $\rho_{AB}$. According PPT
criterion, if $\rho^{T_B}$ has non negative eigenvalues, then $\rho_{AB}$ is
separable.
\begin{lstlisting}
julia> ppt(ρ ⊗ σ, [2, 2], 2)
 0.052512626584708365

julia> ppt(proj((1/sqrt(2)*(ket(1,2) ⊗ ket(1,2)-ket(2,2) ⊗ ket(2,2)))), [2, 2], 2)
-0.4999999999999999
\end{lstlisting}

Another way to quantification of quantum entanglement is \emph{Concurrence}
\cite{hill1997entanglement}. Concurrence of quantum state $\rho$ is a strong
separability criterion. For two-qubit systems it is defined as
$C(\rho)=\max(0,\lambda_1-\lambda_2-\lambda_3-\lambda_4)$, where $\lambda_i$ are
decreasing eigenvalues of $\sqrt{\sqrt{\rho}\tilde{\rho}\sqrt{\rho}}$ with
$\tilde{\rho}=(\sigma_y\otimes\sigma_y)\rho^*(\sigma_y\otimes\sigma_y)$. If 
$C(\rho)=0$, then $\rho$ is separable.

\begin{lstlisting}
julia> ρ = [0.25 0.1im; -0.1im 0.75]
2×2 Array{Complex{Float64},2}:
 0.25+0.0im   0.0+0.1im
-0.0-0.1im  0.75+0.0im

julia> σ = [0.4 0.1im; -0.1im 0.6]
2×2 Array{Complex{Float64},2}:
 0.4+0.0im  0.0+0.1im
-0.0-0.1im  0.6+0.0im

julia> concurrence(ρ ⊗ σ)
 0.0

julia> concurrence(proj(max_entangled(4)))
 0.9999999999999998
\end{lstlisting}

\section*{Measurements}
Measurements are modeled in two ways:
\begin{itemize}
\item as Positive Operator Valued Measures (POVMs),
\item measurements with post-selection.
\end{itemize}
In both cases a~measurement is treated as a special case of a quantum channel
(operation).

\subsection*{Positive Operator Valued Measure measurement} A POVM measurement is
defined as follows. Let $\mu:\Gamma\to\mathrm{P}(\mathcal{X})$ be a mapping from
a finite alphabet of measurement outcomes to the set of linear positive
operators. If $\sum_{\xi\in\Gamma} {\mu(\xi)=\mathbb{I}_{\mathcal{X}}}$ then
$\mu$ is a POVM measurement. The set of positive semi-definite linear operators
is defined as $\mathrm{P}(\mathcal{X})=\{X\in \mathrm{L}(\mathcal{X}):
\bra{\psi}X\ket{\psi}\geq 0 \text{ for all } \ket{\psi}\in\mathcal{X}\}$. POVM
measurement models the situation where a quantum object is destroyed during the
measurement process and quantum state after the measurement does not exists.

We model POVM measurement as a channel
$\theta:\mathrm{L}(\mathcal{X})\to \mathrm{L}(\mathcal{Y})$, where
$\mathcal{Y}=\mathrm{span}\{\ket{\xi}\}_{\xi\in\Gamma}$ such that $\theta(\rho)
= \sum_{\xi\in\Gamma} \tr(\rho\, \mu(\xi))\ketbra{\xi}$. This channel transforms
the measured quantum state into a~classical state (diagonal matrix) containing
probabilities of measuring given outcomes. Note that in \texttt{QuantumInformation.jl}
$\Gamma=\{1,2,\ldots,|\Gamma|\}$ and POVM measurements are represented by the
type 
\begin{lstlisting}
POVMMeasurement{T} <: AbstractQuantumOperation{T} where
T<:AbstractMatrix{<:Number}
\end{lstlisting}
Predicate function \lstinline|ispovm()| verifies whether a~list of matrices is a
proper POVM.

\begin{lstlisting}
julia> ρ=proj(1.0/sqrt(2)*(ket(1,3)+ket(3,3)))
3×3 Array{Complex{Float64},2}:
 0.5+0.0im  0.0+0.0im  0.5+0.0im
 0.0+0.0im  0.0+0.0im  0.0+0.0im
 0.5+0.0im  0.0+0.0im  0.5+0.0im

julia> E0 = proj(ket(1,3))
3×3 Array{Complex{Float64},2}:
 1.0+0.0im  0.0+0.0im  0.0+0.0im
 0.0+0.0im  0.0+0.0im  0.0+0.0im
 0.0+0.0im  0.0+0.0im  0.0+0.0im

julia> E1 = proj(ket(2,3))+proj(ket(3,3))
3×3 Array{Complex{Float64},2}:
 0.0+0.0im  0.0+0.0im  0.0+0.0im
 0.0+0.0im  1.0+0.0im  0.0+0.0im
 0.0+0.0im  0.0+0.0im  1.0+0.0im

julia> M = POVMMeasurement([E0,E1])
POVMMeasurement{Array{Complex{Float64},2}}
dimensions: (3, 2)
Complex{Float64}
 [1.0+0.0im 0.0+0.0im 0.0+0.0im; 
 0.0+0.0im 0.0+0.0im 0.0+0.0im; 
 0.0+0.0im 0.0+0.0im 0.0+0.0im]
Complex{Float64}
 [0.0+0.0im 0.0+0.0im 0.0+0.0im; 
 0.0+0.0im 1.0+0.0im 0.0+0.0im; 
 0.0+0.0im 0.0+0.0im 1.0+0.0im]

julia> ispovm(M)
 true

julia> M(ρ)
2×2 LinearAlgebra.Diagonal{Float64,Array{Float64,1}}:
 0.5  .
 .   0.5
\end{lstlisting}

\subsection*{Measurement with post-selection} When a quantum system after being
measured is not destroyed one can be interested in its state after the
measurement. This state depends on the measurement outcome. In this case the
measurement process is defined in the following way.

Let $\mu:\Gamma\to \Lrm(\mathcal{X}, \mathcal{Y})$ be a mapping from a finite
set of measurement outcomes to set of linear operators called effects. If
$\sum_{\xi\in\Gamma} {\mu(\xi)^\dagger \mu(\xi)=\mathbb{I}_{\mathcal{X}}}$ then
$\mu$ is a quantum measurement. Given outcome $\xi$ was obtained, the state
before the measurement, $\rho$, is transformed into sub-normalized quantum state
$\rho_\xi=\mu(\xi)\rho\mu(\xi)^\dagger$. The outcome $\xi$ will be obtained with
probability $\tr(\rho_\xi)$.
\begin{lstlisting}
julia> PM = PostSelectionMeasurement(E1)
PostSelectionMeasurement{Array{Complex{Float64},2}}
dimensions: (3, 3)
Complex{Float64}
 [0.0+0.0im 0.0+0.0im 0.0+0.0im; 
 0.0+0.0im 1.0+0.0im 0.0+0.0im; 
 0.0+0.0im 0.0+0.0im 1.0+0.0im]

julia> iseffect(PM)
 true

julia> PM(ρ)
3×3 Array{Complex{Float64},2}:
 0.0+0.0im  0.0+0.0im  0.0+0.0im
 0.0+0.0im  0.0+0.0im  0.0+0.0im
 0.0+0.0im  0.0+0.0im  0.5+0.0im
\end{lstlisting}
In \texttt{QuantumInformation.jl} this kind of measurement is modeled as CP-TNI map with a
single Kraus operator $\mu(\xi)$ and represented as
\begin{lstlisting}
PostSelectionMeasurement{T} <: AbstractQuantumOperation{T} where
T<:AbstractMatrix{<:Number}
\end{lstlisting}

Measurement types can be composed and converted to Kraus operators,
superoperators, Stinespring representation operators, and dynamical matrices.
\begin{lstlisting}
julia> α = 0.3
 0.3

julia> K0 = ComplexF64[0 0 sqrt(α); 0 1 0; 0 0 0]
3×3 Array{Complex{Float64},2}:
 0.0+0.0im  0.0+0.0im  0.547723+0.0im
 0.0+0.0im  1.0+0.0im       0.0+0.0im
 0.0+0.0im  0.0+0.0im       0.0+0.0im

julia> K1 = ComplexF64[1 0 0; 0 0 0; 0 0 sqrt(1 - α)]
3×3 Array{Complex{Float64},2}:
 1.0+0.0im  0.0+0.0im      0.0+0.0im
 0.0+0.0im  0.0+0.0im      0.0+0.0im
 0.0+0.0im  0.0+0.0im  0.83666+0.0im

julia> Φ = KrausOperators([K0,K1])
KrausOperators{Array{Complex{Float64},2}}
dimensions: (3, 3)
Complex{Float64}
 [0.0+0.0im 0.0+0.0im 0.547723+0.0im; 
 0.0+0.0im 1.0+0.0im 0.0+0.0im; 
 0.0+0.0im 0.0+0.0im 0.0+0.0im]
Complex{Float64}
 [1.0+0.0im 0.0+0.0im 0.0+0.0im; 
 0.0+0.0im 0.0+0.0im 0.0+0.0im; 
 0.0+0.0im 0.0+0.0im 0.83666+0.0im]

julia> ρ=proj(1.0/sqrt(2)*(ket(1,3)+ket(3,3)))
3×3 Array{Complex{Float64},2}:
 0.5+0.0im  0.0+0.0im  0.5+0.0im
 0.0+0.0im  0.0+0.0im  0.0+0.0im
 0.5+0.0im  0.0+0.0im  0.5+0.0im

julia> (PM ∘ Φ)(ρ)
3×3 Array{Complex{Float64},2}:
 0.0+0.0im  0.0+0.0im   0.0+0.0im
 0.0+0.0im  0.0+0.0im   0.0+0.0im
 0.0+0.0im  0.0+0.0im  0.35+0.0im
\end{lstlisting}

\section*{Random quantum objects}
In this section we present the implementation of the sub-package
\texttt{RandomMatrices}. The justification for including these functionalities
in our package is twofold. First, the application of random matrix theory (RMT)
in quantum information is a blooming field of research with a plethora of
interesting
results~\cite{collins2016random,wootters1990random,zyczkowski2001induced,%
sommers2004statistical,puchala2016distinguishability,zhang2017average,%
zhang2017average2,bruzda2009random,nechita2018almost,zhang2018spectral}.
Hence, it is useful to have readily available implementations of known
algorithms of generating random matrices. Secondly, when performing numerical
investigations, we often need ``generic'' inputs. Generating random matrices
with a known distribution is one of the ways to obtain such generic inputs.

\subsection*{Ginibre matrices}

In this section we introduce the Ginibre random matrices
ensemble~\cite{ginibre1965statistical}. This ensemble is at the core of a vast
majority of algorithms for generating random matrices presented in later
subsections. Let $(G_{ij})_{1 \leq i \leq m, 1 \leq j \leq n}$ be a $m\times n$
table of independent identically distributed (i.i.d.) random variable on
$\mathbb{K}$. The field $\mathbb{K}$ can be either of $\mathbb{R}$, $\C$ or
$\mathbb{Q}$. With each of the fields we associate a Dyson index $\beta$ equal
to $1$, $2$, or $4$ respectively. Let $G_{ij}$ be i.i.d random variables with
the real and imaginary parts sampled independently from the distribution
$\mathcal{N}(0, \frac{1}{\beta})$. Hence, $G \in \Lrm(\XX, \YY)$, where
matrix $G$ is
\begin{equation}
P(G) \propto \exp(-\Tr G G^\dagger).
\end{equation}
This law is unitarily invariant, meaning that for any unitary matrices $U$ and
$V$, $G$ and $UGV$ are equally distributed. It can be shown that for $\beta=2$
the eigenvalues of $G$ are uniformly distributed over the unit disk on the
complex plane~\cite{tao2008random}.

In our library the ensemble Ginibre matrices is implemented in the
\lstinline|GinibreEnsemble{β}| parametric type. The parameter determines the
Dyson index. The following constructors are provided
\begin{lstlisting}
julia> GinibreEnsemble{β}(m::Int, n::Int)
julia> GinibreEnsemble{β}(m::Int)
julia> GinibreEnsemble(m::Int, n::Int)
julia> GinibreEnsemble(m::Int)
\end{lstlisting}
The parameters $n$ and $m$ determine the dimensions of output and input spaces.
The versions with one argument assume $m=n$. When the Dyson index is omitted
it assumed that $\beta=2$. Sampling from these distributions can be performed as
follows
\begin{lstlisting}
julia> g = GinibreEnsemble{2}(2,3)
 GinibreEnsemble{2}(m=2, n=3)

julia> rand(g)
2×3 Array{Complex{Float64},2}:
 0.835803+1.10758im   -0.622744-0.130165im  -0.677944+0.636562im
 1.32826+0.106582im   -0.460737-0.531975im  -0.656758+0.0244259im
\end{lstlisting}
The function \lstinline|rand| has specialized methods for each possible value
of the Dyson index $\beta$.

\subsection*{Wishart matrices}

Wishart matrices form an ensemble of random positive semidefinite matrices. They
are parametrized by two factors. First is the Dyson index $\beta$ which is equal
to one for real matrices, two for complex matrices and four for symplectic
matrices. The second parameter, $K$, is responsible for the rank of the
matrices. They are sampled as follows
\begin{enumerate}
\item Choose $\beta$ and $K$.

\item Sample a Ginibre matrix $G\in \Lrm(\XX, \YY)$ with the Dyson index $\beta$
and $\mathrm{dim}(\XX) = d$ and $\mathrm{dim}(\YY)=Kd$.

\item Return $GG^\dagger$.
\end{enumerate}

In \texttt{QuantumInformation.jl} this is implemented using the type
\lstinline|WishartEnsemble{β, K}|. We also provide additional constructors for
convenience
\begin{lstlisting}
 WishartEnsemble{β}(d::Int) where β = WishartEnsemble{β, 1}(d)
 WishartEnsemble(d::Int) = WishartEnsemble{2}(d)
\end{lstlisting}

These can be used in the following way
\begin{lstlisting}
julia> w = WishartEnsemble{1,0.2}(5)
 WishartEnsemble{1,0.2}(d=5)

julia> z = rand(w)
5×5 Array{Float64,2}:
 0.0897637  0.0257443   0.0314593   0.0223569   0.093517
 0.0257443  0.00738347  0.00902253  0.00641196  0.0268207
 0.0314593  0.00902253  0.0110254   0.00783535  0.0327746
 0.0223569  0.00641196  0.00783535  0.00556828  0.0232917
 0.093517   0.0268207   0.0327746   0.0232917   0.0974271

julia> eigvals(z)
5-element Array{Float64,1}:
-1.549149323294561e-17
-1.11670454111383e-18
 1.5797866551971292e-18
 6.408793727745745e-18
 0.21116803949130986

julia> w = WishartEnsemble(3)
 WishartEnsemble{2,1}(d=3)

julia> z = rand(w)
3×3 Array{Complex{Float64},2}:
 0.474628+0.0im         0.177244-0.0227445im   0.137337-0.0929298im
 0.177244+0.0227445im   0.128676+0.0im         0.0938587-0.165916im
 0.137337+0.0929298im   0.0938587+0.165916im   0.555453+0.0im  

julia> eigvals(z)
3-element Array{Float64,1}:
 0.01707438064450695
 0.35884924300093163
 0.7828337014291611
\end{lstlisting}
\subsection*{Circular ensembles}
Circular ensembles are measures on the space of unitary matrices. There are
three main circular ensembles. Each of this ensembles has an associated Dyson
index $\beta$~\cite{mehta2004random}
\begin{itemize}
\item Circular orthogonal ensemble (COE), $\beta=1$.
\item Circular unitary ensemble (CUE), $\beta=2$.
\item Circular symplectic ensemble (CSE), $\beta=4$.
\end{itemize}
They can be characterized as follows. The CUE is simply the Haar measure on the
unitary group. Now, if $U$ is an element of CUE then $U^TU$ is an element of
$COE$ and $U_R U$ is an element CSE. Here
\begin{equation}
U_R = \left( \begin{array}{ccccccc} 0 & -1 & & & & &  \\ 1 & 0 &  & & & & \\ &
& 0 & -1 &  & &  \\ & & 1 & 0  & & & \\ & & & & \ddots & & \\ & & & & & 0& -1\\
& & & & & 1 & 0 \end{array} \right) U^T \left( \begin{array}{ccccccc} 0 & 1 & &
& & &  \\ -1 & 0 &  & & & & \\ & & 0 & 1 &  & &  \\ & & -1 & 0  & & & \\ & & &
& \ddots & & \\ & & & & & 0& 1\\ & & & & & -1 & 0 \end{array} \right).
\end{equation}
As can be seen the sampling of Haar unitaries is at the core of sampling these
ensembles. Hence, we will focus on them in the remainder of this section.

There are several possible approaches to generating random unitary matrices
according to the Haar measure. One way is to consider known parametrizations of
unitary matrices, such as the Euler~\cite{zyczkowski1994random} or
Jarlskog~\cite{jarlskog2005recursive} ones. Sampling these parameters from
appropriate distributions yields a Haar random unitary. The downside is the long
computation time, especially for large matrices, as this involves a lot of
matrix multiplications. We will not go into this further, we refer the
interested reader to the papers on these parametrizations.

Another approach is to consider a Ginibre matrix $G \in \Lrm(\XX)$ and its polar
decomposition $G=U P$, where $U \in \Lrm(\XX)$ is unitary and $P$ is a positive
matrix. The matrix $P$ is unique and given by $\sqrt{G^\dagger G}$. Hence,
assuming $P$ is invertible, we could recover $U$ as
\begin{equation}
U = G (G^\dagger G) ^{-\frac{1}{2}}.
\end{equation}
As this involves the inverse square root of a matrix, this approach can be
potentially numerically unstable.

The optimal approach is to utilize the QR decomposition of $G$, $G=QR$, where $Q
\in \Lrm(\XX)$ is unitary and $R \in \Lrm(\XX)$ is upper triangular. This
procedure is unique if $G$ is invertible and we require the diagonal elements of
$R$ to be positive. As typical implementations of the QR algorithm do not
consider this restriction, we must enforce it ourselves. The algorithm is as
follows
\begin{enumerate}
\item Generate a Ginibre matrix $G \in \Lrm(\XX)$, $\mathrm{dim}(\XX) = d$ with
Dyson index $\beta=2$.

\item Perform the QR decomposition obtaining $Q$ and $R$.

\item Multiply the $i$\textsuperscript{th} column of $Q$ by $r_{ii}/|r_{ii}|$.
\end{enumerate}
This gives us a Haar distributed random unitary. For detailed analysis of this
algorithm see~\cite{mezzadri2006generate}. This procedure can be generalized in
order to obtain a random isometry. The only required changed is the dimension of
$G$. We simply start with $G \in \Lrm(\XX, \YY)$, where $\dim(\XX)\geq
\dim(\YY)$.

Furthermore, we may introduce two additional circular ensembles corresponding
to the Haar measure on the orthogonal and symplectic groups. These are the
circular real ensemble (CRE) and circular quaternion ensemble (CQE). Their
sampling is similar to sampling from CUE. The only difference is the initial
Dyson index of the Ginibre matrix. This is set to $\beta=1$ for CRE and
$\beta=4$ for CQE.

In \lstinline|QuantumInformation.jl| these distributions can be sampled as
\begin{lstlisting}
julia>  c = CircularEnsemble{2}(3)
CircularEnsemble{2}(
d: 3
g: GinibreEnsemble{2}(m=3, n=3)
)

julia> u = rand(c)
3×3 Array{Complex{Float64},2}:
 0.339685+0.550434im  -0.392266-0.3216im      -0.53172+0.203988im
 0.515118-0.422262im   0.392165-0.626859im    -0.0504431-0.084009im
 0.297203+0.222832im  -0.418737-0.143578im     0.607012-0.545525im


julia> u*u'
3×3 Array{Complex{Float64},2}:
 1.0+0.0im                    -5.55112e-17-5.55112e-17im   -2.77556e-17-4.16334e-17im
-5.55112e-17+5.55112e-17im     1.0+0.0im                   -2.498e-16+0.0im        
-2.77556e-17+4.16334e-17im    -2.498e-16+0.0im              1.0+0.0im  
\end{lstlisting}

Sampling from the Haar measure on the orthogonal group can be achieved as
\begin{lstlisting}
julia> c = CircularRealEnsemble(3)
CircularRealEnsemble(
d: 3
g: GinibreEnsemble{1}(m=3, n=3)
)

julia> o = rand(c)
3×3 Array{Float64,2}:
 0.772464    0.611349  -0.171907
 0.0524376   0.208368   0.976644
 0.63289    -0.763436   0.128899

julia> o*o'
3×3 Array{Float64,2}:
 1.0          -1.38778e-16  -8.67362e-17
-1.38778e-16   1.0           8.32667e-17
-8.67362e-17   8.32667e-17   1.0
\end{lstlisting}

For convenience we provide the following type aliases

\begin{lstlisting}
const COE = CircularEnsemble{1}
const CUE = CircularEnsemble{2}
const CSE = CircularEnsemble{4}
\end{lstlisting}
\subsection*{Random quantum states}

In this section we discuss the properties and methods of generating random
quantum states. We will treat quantum channels as a special case of quantum
states.

\subsubsection*{Pure states}

Pure states are elements of the unit sphere in $\XX$. Thus it is straightforward
to generate them randomly. We start with a vector of $\dim(\XX)$ independent
complex numbers sampled from the standard normal distribution. What remains is
to normalize the length of this vector to unity.

This is implemented using the \lstinline|HaarKet{β}| type. The value $\beta=1$
corresponds to the Haar measure on the unit sphere in $\mathbb{R}^d$, while
$\beta=2$ corresponds to the Haar measure on the unit sphere in $\C^d$. The
usage is as follows

\begin{lstlisting}
julia> h = HaarKet{2}(3)
HaarKet{2}(d=3)

julia> ψ = rand(h)
3-element Array{Complex{Float64},1}:
 0.1687649644765863 - 0.3201009507269653im
 0.7187423269572294 - 0.39405022770434767im
 0.1342475675218075 + 0.42327915636096036im

julia> norm(ψ)
 1.0
\end{lstlisting}
For convenience we provide the following constructor
\begin{lstlisting}
HaarKet(d::Int) = HaarKet{2}(d)
\end{lstlisting}
as the majority of uses cases require sampling complex states.

\subsubsection*{Mixed states}

Random mixed states can be generated in one of two equivalent ways. The first
one comes from the partial trace of random pure states. Suppose we have a pure
state $\ket{\psi} \in \XX \otimes \YY$. Then we can obtain a random mixed as
\begin{equation}
\rho = \tr_\YY \ketbra{\psi}{\psi}.
\end{equation}
Note that in the case $\dim(\XX)=\dim(\YY)$ we recover the (flat)
Hilbert-Schmidt distribution on the set of quantum states.

An alternative approach is to start with a Ginibre matrix $G \in \Lrm(\XX,
\YY)$. We obtain a random quantum state $\rho$ as
\begin{equation}
\rho = GG^\dagger/\Tr(GG^\dagger).
\end{equation}
It can be easily verified that this approach is equivalent to the one utilizing
random pure states. First, note that in both cases we start with $\dim(\XX)
\dim(\YY)$ complex random numbers sampled from the standard normal
distribution. Next, we only need to note that taking the partial trace of a
pure state $\ket{\psi}$ is equivalent to calculating $AA^\dagger$ where $A$ is
a matrix obtained from reshaping $\ket{\psi}$.

The properties of these states have been extensively studied. We will omit
stating all the properties here and refer the reader
to~\cite{wootters1990random,zyczkowski2001induced,
sommers2004statistical,puchala2016distinguishability,zhang2017average,zhang2017average2}.

Sampling random mixed states is implemented using the
\lstinline|HilbertSchmidtStates{β, K}| type. The meaning of the type parameters
is the same as in the Wishart matrices case. We provide additional constructors
which set the default values of the parameters
\begin{lstlisting}
HilbertSchmidtStates{β}(d::Int) where β = HilbertSchmidtStates{β, 1}(d)
HilbertSchmidtStates(d::Int) = HilbertSchmidtStates{2, 1}(d)
\end{lstlisting}
The latter one is the most frequent use case. Here is an example
\begin{lstlisting}
julia> h = HilbertSchmidtStates(3)
HilbertSchmidtStates{2,1}(WishartEnsemble{2,1}(d=3), 3)

julia> ρ = rand(h)
3×3 Array{Complex{Float64},2}:
 0.335603+0.0im           0.0696096+0.0606972im     0.0373103+0.0853966im
 0.0696096-0.0606972im    0.209561+0.0im           -0.000865656+0.0129982im
 0.0373103-0.0853966im   -0.000865656-0.0129982im   0.454836+0.0im

julia> tr(ρ)
 1.0 + 0.0im

julia> eigvals(ρ)
3-element Array{Float64,1}:
 0.15460054248543945
 0.3306739537037592
 0.5147255038108014
\end{lstlisting}
\subsection*{Random quantum channels}

Quantum channels are a special subclass of quantum states with constraints
imposed on their \emph{partial} trace as well as trace. Formally, we start with
a Ginibre matrix $G \in \Lrm (\XX \otimes \YY, \ZZ)$. We obtain a random
Choi-Jamio{\l}kowski matrix $J_\Phi$ corresponding to a channel $\Phi$ as
\begin{equation}
J_\Phi = \left( \mathbb{I}_{\XX} \otimes (\Tr_\XX GG^\dagger)^{-1/2} \right)
GG^\dagger \left( \mathbb{I}_{\XX} \otimes (\Tr_\XX GG^\dagger)^{-1/2} \right).
\end{equation}
When $\dim(\ZZ)=\dim(\XX) \dim(\YY)$ this is known to generate a uniform
distribution over the set of quantum
channels~\cite{bruzda2009random,nechita2018almost}.

The implementation uses the type \lstinline|ChoiJamiolkowskiMatrices{β, K}|. The
parameters $\beta$ and $K$ have the same meaning as in the Wishart matrix case.
Additionally here, the constructor
\begin{lstlisting}
ChoiJamiolkowskiMatrices{β, K}(idim::Int, odim::Int)  where {β, K}
\end{lstlisting}
takes two parameters---the input and output dimension of the channel. As in the
previous cases we provide some additional constructors for convenience
\begin{lstlisting}
function ChoiJamiolkowskiMatrices{β}(idim::Int, odim::Int) where β
    ChoiJamiolkowskiMatrices{β, 1}(idim, odim)
end

function ChoiJamiolkowskiMatrices{β}(d::Int) where β
    ChoiJamiolkowskiMatrices{β}(d, d)
end

function ChoiJamiolkowskiMatrices(idim::Int, odim::Int)
    ChoiJamiolkowskiMatrices{2}(idim, odim)
end

function ChoiJamiolkowskiMatrices(d::Int)
    ChoiJamiolkowskiMatrices(d, d)
end
\end{lstlisting}

Here is an example of usage
\begin{lstlisting}
julia> c = ChoiJamiolkowskiMatrices(2, 3)
ChoiJamiolkowskiMatrices{2,1}(WishartEnsemble{2,1}(d=6), 2, 3)

julia> Φ = rand(c)
DynamicalMatrix{Array{Complex{Float64},2}}
dimensions: (2, 3)
Complex{Float64}
 [0.307971-4.98733e-18im -0.00411588+0.0368471im … 
-0.0676732+0.024328im  0.0860858+0.00302876im; 
-0.00411588-0.0368471im 0.167651+2.1684e-19im … 
-0.0428561+0.0266119im    0.0191888+0.0101013im; 
 … ; 
-0.0676732-0.024328im -0.0428561-0.0266119im … 
 0.210419+0.0im -0.103401-0.142753im; 
 0.0860858-0.00302876im 0.0191888-0.0101013im … 
-0.103401+0.142753im 0.411068+0.0im]

julia> ptrace(Φ.matrix, [3, 2],[1])
2×2 Array{Complex{Float64},2}:
 1.0-1.53957e-17im             -1.38778e-17-3.05311e-16im
 1.38778e-17+3.05311e-16im      1.0+2.1684e-19im
\end{lstlisting}
Note that the resulting sample is of type \lstinline|DynamicalMatrix|.

\section*{Example} 
\begin{figure}[!h]
\includegraphics[width=\textwidth]{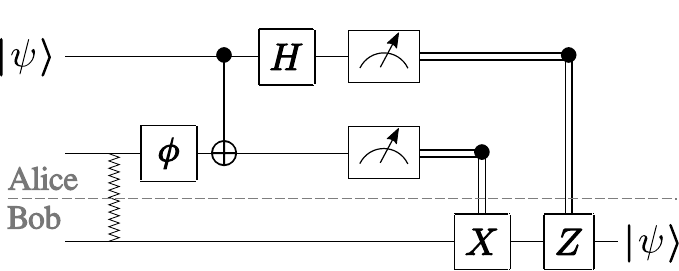}
\caption{Schematic depiction of the teleportation protocol. 
Squiggly line represents the maximally entangled state, 
and box $\Phi$ represents the noise operator.}\label{fig:teleportiation}
\end{figure}

As an example we provide the teleportation protocol in the
presence of noise. Imagine we have an entangled pair of particles in the state 
\begin{equation}
\ket{\psi} = \frac{1}{\sqrt{2}} \left( \ket{00} + \ket{11} \right).
\end{equation}
One of the particles stays with Alice and another is sent through a noisy
channel to Bob. As a noise model we chose the amplitude damping channel given by
the Kraus operators
\begin{equation}
K_0 = \begin{pmatrix}
1 & 0 \\
0 & \sqrt{1-\gamma}
\end{pmatrix} \quad
K_1 = \begin{pmatrix}
0 & \sqrt{\gamma} \\
0 & 0
\end{pmatrix}
.
\end{equation}
The channel has one parameter $\gamma \in [0, 1]$ modeling the strength of the
noise. Assume that Alice possesses a random pure state $\ket{\phi}$ that she
teleports to Bob. The protocol is shown in Fig.~\ref{fig:teleportiation}.

Our examples show the fidelity of the final state at Bob's site averaged over
$100$ random pure initial states. We also check how the parameter $\gamma$
influences this fidelity.
\begin{lstlisting}
using QuantumInformation

steps = 100
haar = HaarKet(2)
ψ = (ket(0, 4) + ket(3, 4))/sqrt(2)
γs = 0.0:0.01:1.0
Φ = KrausOperators([[1 0; 0 sqrt(1-γ)], [0 sqrt(γ); 0 0]])
post = [PostSelectionMeasurement(proj(ket(i, 4)) ⊗ eye(2)) for i=0:3]
rots = [UnitaryChannel(eye(2)), UnitaryChannel(sx), UnitaryChannel(sz),
UnitaryChannel(sx*sz)]
had = UnitaryChannel{Matrix{ComplexF64}}(hadamard(2))
cnot = UnitaryChannel{Matrix{ComplexF64}}([1 0 0 0; 0 1 0 0; 0 0 0 1; 0 0 1 0])
r = zeros(steps, length(γs), 4);
for (k, γ) in enumerate(γs)
    for i=1:steps
        ϕ = rand(haar)
        ξ = ϕ ⊗ ψ
        ρ = ((had ⊗ IdentityChannel(4))∘(cnot ⊗
IdentityChannel(2))∘(IdentityChannel(4) ⊗ Φ))(ξ)
        for j=1:4
            σ = rots[j](ptrace(post[j](ρ), [2, 2, 2], [1, 2]))
            r[i, k, j] = fidelity(ϕ, σ/tr(σ))
        end
    end
end
mean(r, 1)
\end{lstlisting}

\section*{Benchmarks}
In the benchmarks we compare our library to the state-of-the-art \texttt{Python}
library, \texttt{QuTiP} \cite{johansson2012qutip,johansson2013qutip}. We perform the following tests:
\begin{enumerate}
\item sampling a random unitary matrix, 
\item sampling a random pure state,
\item sampling a random mixed state,
\item sampling a random channel,
\item calculating the trace distance of a random mixed state from the maximally 
mixed state,
\item calculating the trace distance between two random mixed states,
\item calculating the entropy of the stationary state of a random channel.
\end{enumerate}
The latter is done as follows. First we sample a random quantum channel. Next, 
we apply the \texttt{reshuffle} operation and calculate its eigenvectors and 
eigenvalues. We take the state corresponding to the eigenvalue equal to one and 
calculate its von Neumann entropy. All tests are performed 1000 times and an 
average time of computation from these samples is calculated. In the case of 
\texttt{Julia} code we ensure all functions are compiled prior to testing. The 
tests are performed for dimensions $4$, $16$, $64$, $256$, $1024$. In the next 
subsections we present and discuss results for all the aforementioned cases.

The tests were performed on a machine equipped with Intel\textregistered\
Core\texttrademark\ i7-6800K and 64 GB of RAM. The libraries installed on the
system were
\begin{lstlisting}
julia> versioninfo()
Julia Version 1.0.0
Platform Info:
  OS: Linux (x86_64-redhat-linux)
  CPU: Intel(R) Core(TM) i7-6800K CPU @ 3.40GHz
  WORD_SIZE: 64
  LIBM: libopenlibm
  LLVM: libLLVM-6.0.0 (ORCJIT, broadwell)

julia> LAPACK.version()
v"3.7.0"

julia> BLAS.vendor()
:openblas

julia> BLAS.openblas_get_config()
"DYNAMIC_ARCH NO_AFFINITY Haswell"
\end{lstlisting}
The \texttt{Python} libraries were
\begin{lstlisting}
In [1]: import numpy as np

In [2]: np.__version__
Out[2]: '1.7.1'

In [3]: np.__config__.show()
Out[3]: 
blas_mkl_info:
  NOT AVAILABLE
blis_info:
  NOT AVAILABLE
openblas_info:
    libraries = ['openblas', 'openblas']
    library_dirs = ['/home/user/anaconda3/lib']
    language = c
    define_macros = [('HAVE_CBLAS', None)]
blas_opt_info:
    libraries = ['openblas', 'openblas']
    library_dirs = ['/home/user/anaconda3/lib']
    language = c
    define_macros = [('HAVE_CBLAS', None)]
lapack_mkl_info:
  NOT AVAILABLE
openblas_lapack_info:
    libraries = ['openblas', 'openblas']
    library_dirs = ['/home/user/anaconda3/lib']
    language = c
    define_macros = [('HAVE_CBLAS', None)]
lapack_opt_info:
    libraries = ['openblas', 'openblas']
    library_dirs = ['/home/user/anaconda3/lib']
    language = c
    define_macros = [('HAVE_CBLAS', None)]
\end{lstlisting}

\subsection*{Sampling a random unitary matrix}
\begin{figure}[!h]
\centering\includegraphics{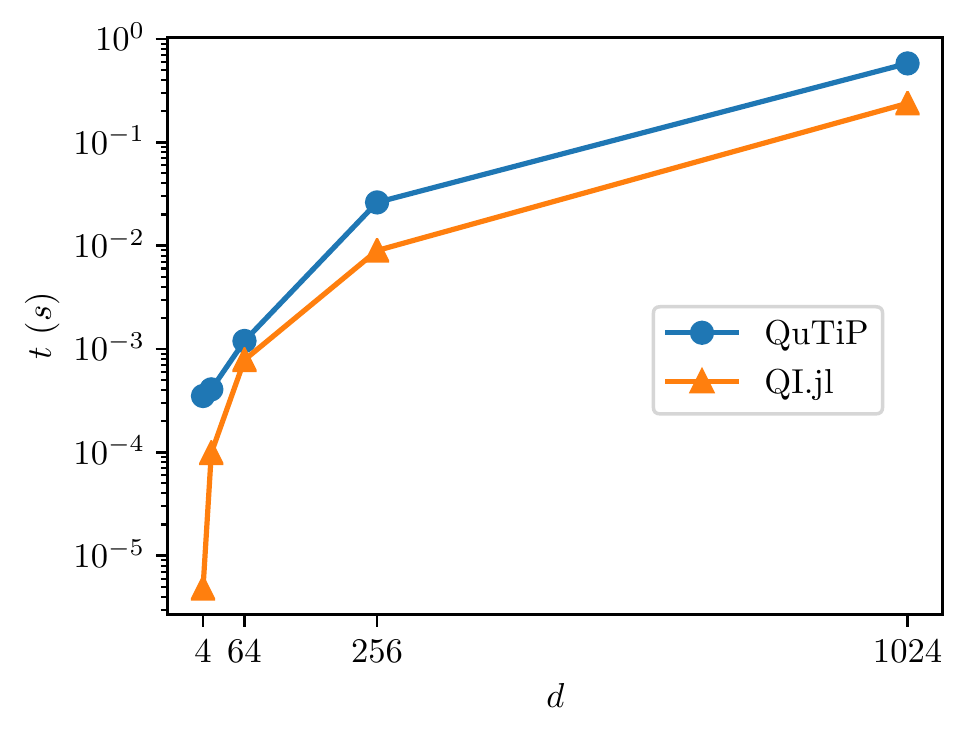}
\caption{Benchmark results for sampling random unitary matrices in 
\texttt{QuantumInformation.jl} and \texttt{Python}. }\label{fig:bench-rand-u}
\end{figure}

The \texttt{Julia} code for this test is
\begin{lstlisting}
using QuantumInformation
function random_unitary(steps::Int, d::Int)
  dist = CUE(d)
  for i=1:steps U = rand(dist) end
end
\end{lstlisting}
The \texttt{Python} implementation reads
\begin{lstlisting}
import qutip as q
def random_unitary(steps, d):
    for _ in range(steps):
        q.rand_unitary_haar(d)
\end{lstlisting}
The benchmark results are presented in Fig.~\ref{fig:bench-rand-u}. Note that, 
as advertised, our implementation is faster and the gap gets bigger as the 
dimension of the input system increases.

\subsection*{Sampling a random pure state}
\begin{figure}[!h]
\centering\includegraphics{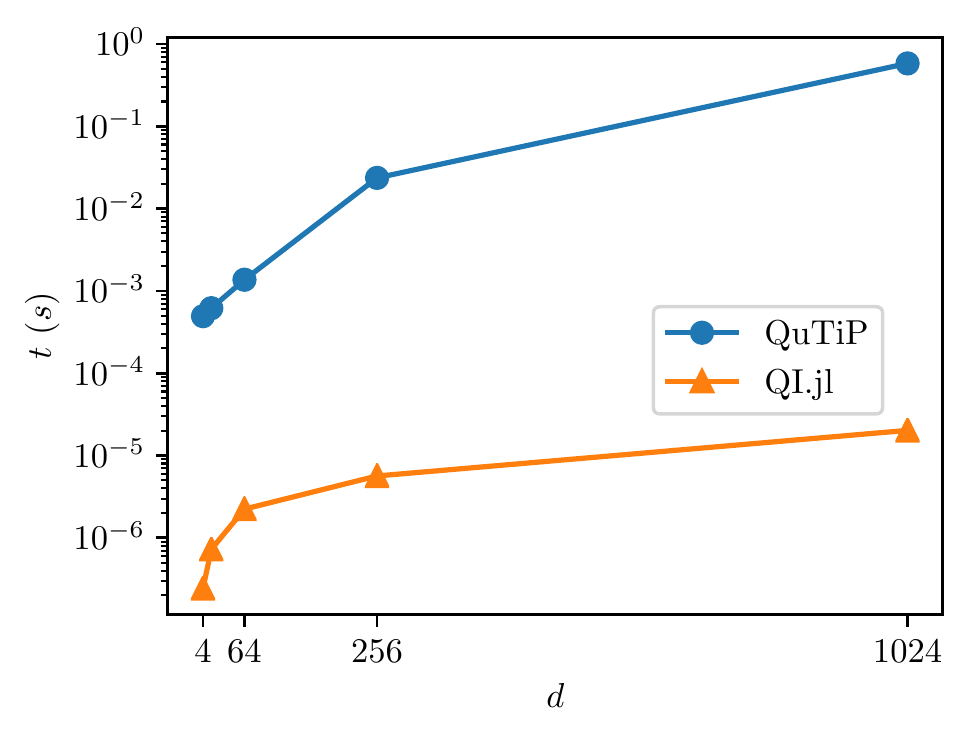}
\caption{Benchmark results for sampling random pure states in 
\texttt{QuantumInformation.jl} and \texttt{Python}.}\label{fig:bench-rand-p}
\end{figure}

The \texttt{Julia} code for this test is
\begin{lstlisting}
using QuantumInformation
function random_pure_state(steps::Int, d::Int)
  dist = HaarKet(d)
  for i=1:steps ψ = rand(dist) end
end
\end{lstlisting}
The \texttt{Python} implementation reads
\begin{lstlisting}
import qutip as q
def random_pure_state(steps, d):
    for _ in range(steps):
        q.rand_ket_haar(d)
\end{lstlisting}
The benchmark results are presented in Fig.~\ref{fig:bench-rand-p}. In this 
case we get a huge difference in the computation times. This is due to the fact 
that \texttt{QuTiP} first samples an entire random unitary matrix and returns 
its first column as the sampled state. On the other hand our implementation 
samples only one vector.

\subsection*{Sampling a random mixed state}
\begin{figure}[!h]
\centering\includegraphics{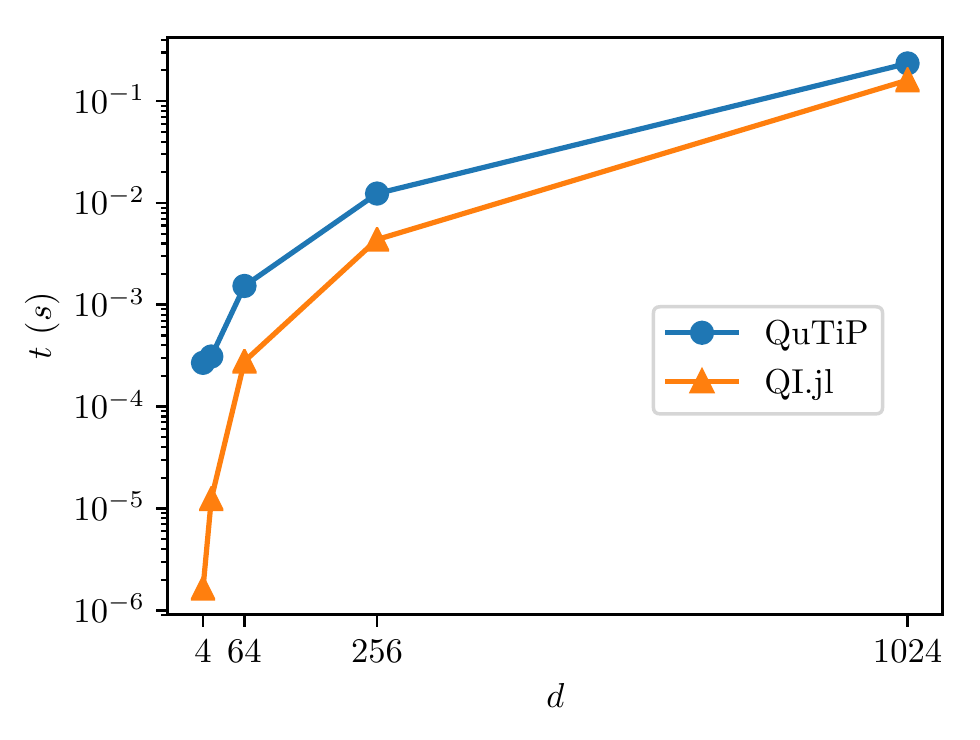}
\caption{Benchmark results for sampling random mixed state in 
\texttt{QuantumInformation.jl} and \texttt{Python}.}\label{fig:bench-rand-m}
\end{figure}

The \texttt{Julia} code for this test is
\begin{lstlisting}
using QuantumInformation
function random_mixed_state(steps::Int, d::Int)
  dist = HilbertSchmidtStates(d)
  for i=1:steps ρ =rand(dist) end
end
\end{lstlisting}
The \texttt{Python} implementation reads
\begin{lstlisting}
import qutip as q
def random_mixed_state(steps, d):
    for _ in range(steps):
        q.rand_dm_hs(d)
\end{lstlisting}
The benchmark results are presented in Fig.~\ref{fig:bench-rand-m}. Again, our 
package is faster compared to \texttt{QuTiP}.

\subsection*{Sampling a random channel}
\begin{figure}[!h]
\centering\includegraphics{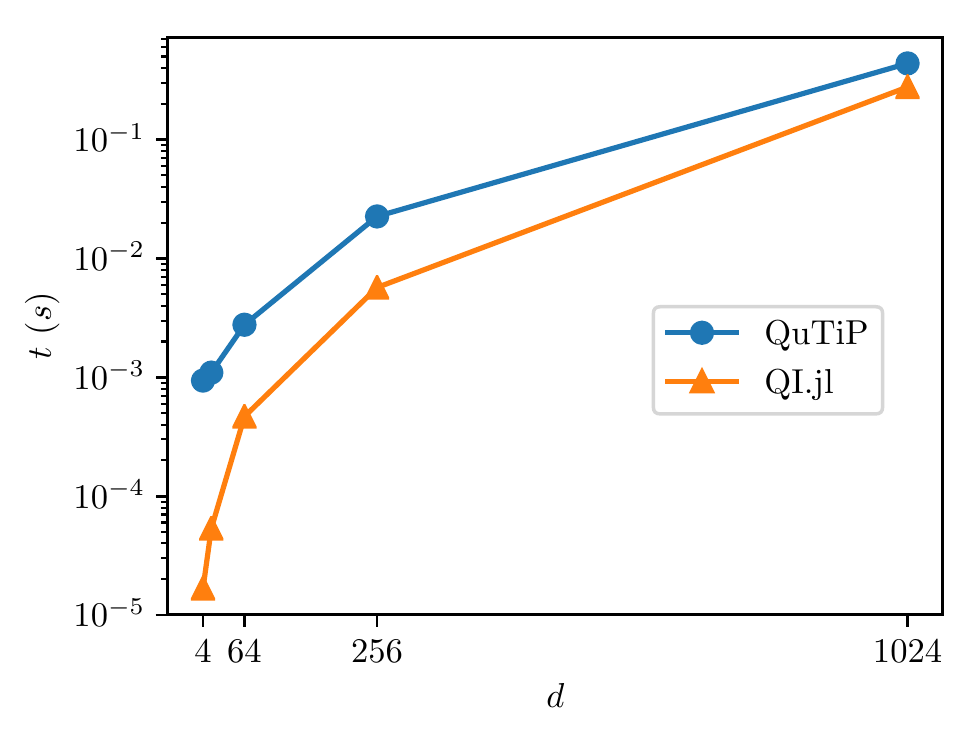}
\caption{Benchmark results for sampling random quantum channels in 
\texttt{QuantumInformation.jl} and \texttt{Python}. }\label{fig:bench-rand-ch}
\end{figure}

The \texttt{Julia} code for this test is
\begin{lstlisting}
using QuantumInformation
function random_channel(steps::Int, d::Int)
  dist = ChoiJamiolkowskiMatrices(round(Int, sqrt(d)))
  for i=1:steps Φ = convert(SuperOperator{Matrix{ComplexF64}}, rand(dist)) end
end
\end{lstlisting}
The \texttt{Python} implementation reads
\begin{lstlisting}
import qutip as q
def random_channel(steps, d):
    for _ in range(steps):
        q.rand_super_bcsz(int(np.sqrt(d)))
\end{lstlisting}
Note the conversion to \texttt{SuperOperator} in the benchmark. This is to mimic
\texttt{QuTiP}'s behavior which returns a superoperator. The benchmark results
are presented in Fig.~\ref{fig:bench-rand-ch}.

\subsection*{Calculating the trace distance form the maximally mixed state}
\begin{figure}[!h]
\centering\includegraphics{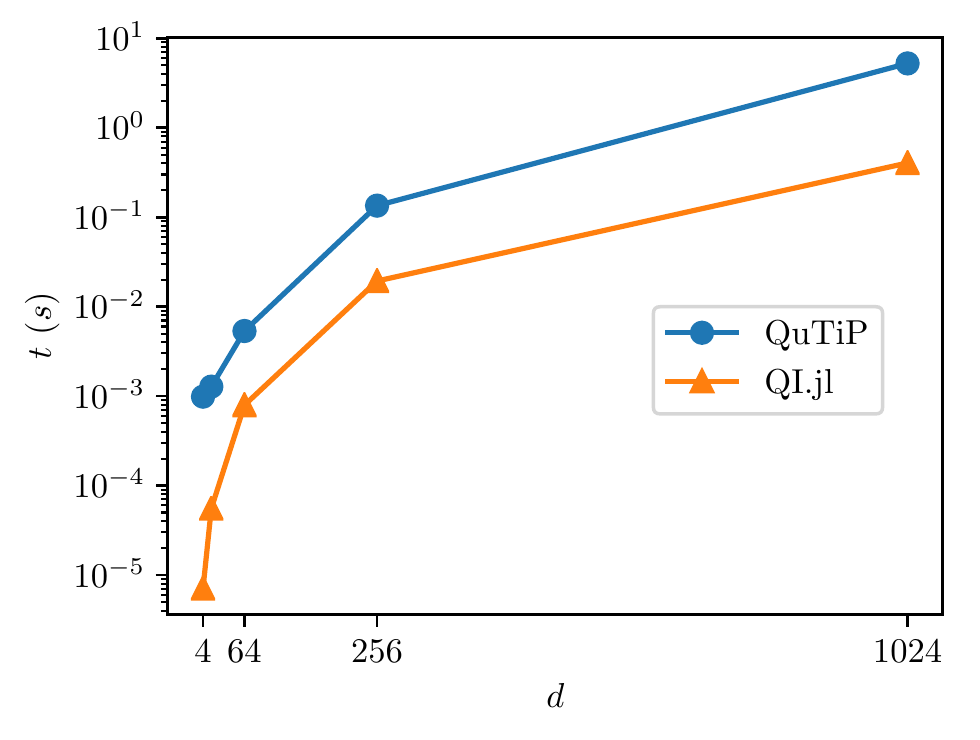}

\caption{Benchmark results for calculating the trace distance between a random
mixed state and the maximally mixed in \texttt{QuantumInformation.jl} and
\texttt{Python}.}\label{fig:bench-rand-trid}
\end{figure}

The \texttt{Julia} code for this test is
\begin{lstlisting}
using QuantumInformation
function trace_distance_max_mixed(steps::Int, d::Int)
  dist = HilbertSchmidtStates(d)
  ρ = 𝕀(d)/d
  for i=1:steps trace_distance(rand(dist), ρ) end
end
\end{lstlisting}
The \texttt{Python} implementation reads
\begin{lstlisting}
import qutip as q
def trace_distance_max_mixed(steps, d):
    rho = q.Qobj(np.eye(d) / d)
    for _ in range(steps):
        q.metrics.tracedist(q.rand_dm_hs(d), rho)
\end{lstlisting}
The benchmark results are presented in Fig.~\ref{fig:bench-rand-trid}. Again, 
for all studied dimensions, our implementation is faster compared to 
\texttt{Python}.

\subsection*{Calculating the trace distance between two random mixed states}
\begin{figure}[!h]
\centering\includegraphics{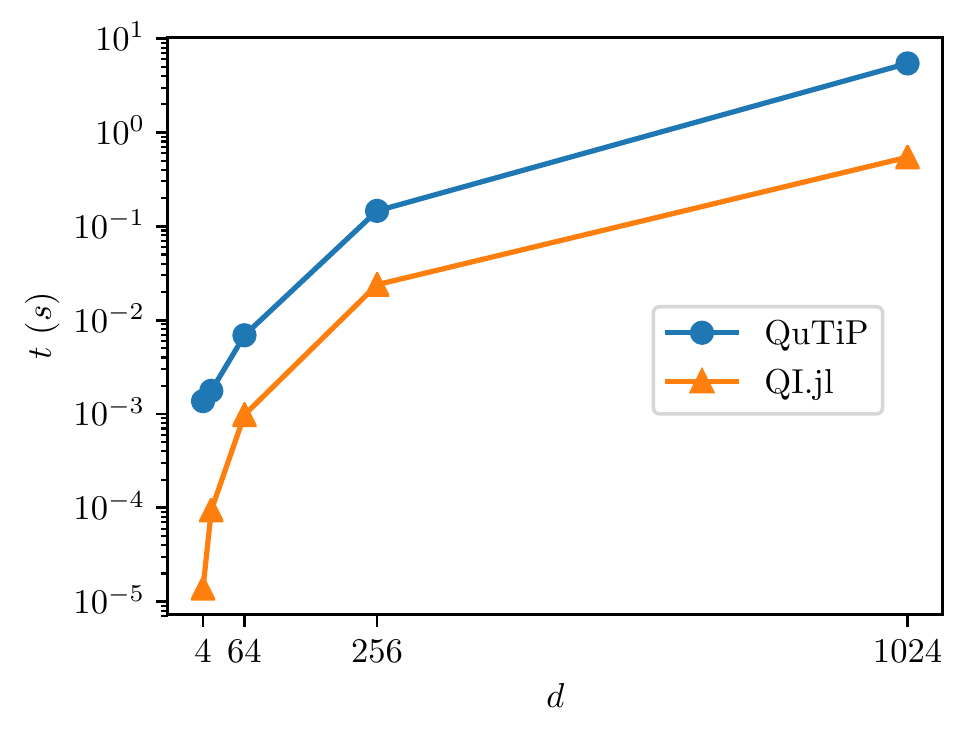}

\caption{Benchmark results for calculating the trace distance between two random
mixed states in \texttt{QuantumInformation.jl} and
\texttt{Python}.}\label{fig:bench-rand-trrand}
\end{figure}

The \texttt{Julia} code for this test is
\begin{lstlisting}
using QuantumInformation
function trace_distance_random(steps::Int, d::Int)
  dist = HilbertSchmidtStates(d)
  for i=1:steps trace_distance(rand(dist), rand(dist)) end
end
\end{lstlisting}
The \texttt{Python} implementation reads
\begin{lstlisting}
import qutip as q
def trace_distance_random(steps, d):
    for _ in range(steps):
        q.metrics.tracedist(q.rand_dm_hs(d), q.rand_dm_hs(d))
\end{lstlisting}
The benchmark results are presented in Fig.~\ref{fig:bench-rand-trrand}.

\subsection*{Calculating the entropy of the stationary state of a random 
channel}
\begin{figure}[!h]
\centering\includegraphics{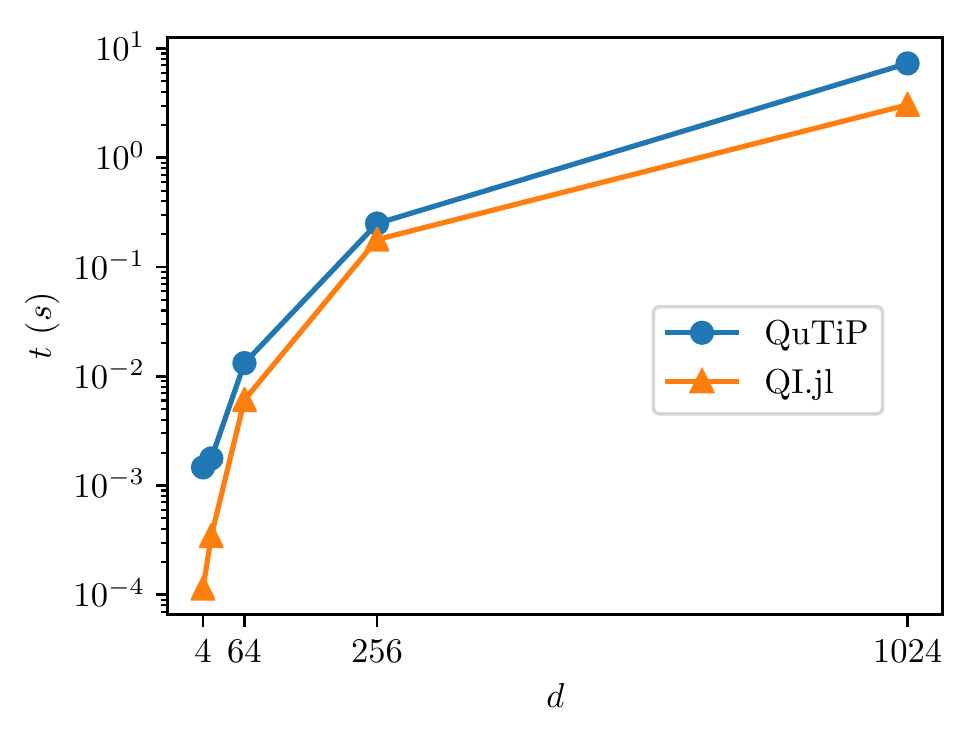}
\caption{Benchmark results for calculating the entropy of a stationary state of 
a random quantum channel in 
\texttt{QuantumInformation.jl} and \texttt{Python}.}\label{fig:bench-rand-entr}
\end{figure}

The \texttt{Julia} code for this test is
\begin{lstlisting}
using QuantumInformation
function random_unitary(steps::Int, d::Int)
  dist = CUE(d)
  for i=1:steps U = rand(dist) end
end
\end{lstlisting}
The \texttt{Python} implementation reads
\begin{lstlisting}
import qutip as q
def random_unitary(steps, d):
    for _ in range(steps):
        q.rand_unitary_haar(d)
\end{lstlisting}
The benchmark results are presented in Fig.~\ref{fig:bench-rand-entr}.

\section*{Conclusions and future work} Numerical investigations are important
part of research in many fields of science, especially in quantum information.
The \texttt{Julia} language is a modern programming language, which provides
strong support for linear algebra and posses an extensive type system. One of
the important feature of \texttt{Julia} is high performance approaching
statically-compiled languages like \texttt{C} or \texttt{Fortran}. Those were
the reasons why we created the \texttt{QuantumInformation.jl} library in 
\texttt{Julia}.

We performed benchmark comparisons of \texttt{QuantumInformation.jl} with \texttt{QuTiP}. They 
clearly state that our library is faster compared to the current state of the 
art. As the core numerical libraries were the same for both tested packages, we 
conclude that this speedup is due to the advantages offered by \texttt{Julia}.

Future work will consists of optimization of numerical code, extending the type
system, developing further functionals, better integration with
\texttt{Convex.jl} package. Additional work will also include parallelization of
the code and support for writing quantum circuits in more intuitive manner.

\section*{Acknowledgments} The authors acknowledge the support of the Polish
National Science Centre. PG under project number 2014/15/B/ST6/05204, DK under
project number 2016/22/E/ST6/00062 and {\L}P under project number
2015/17/B/ST6/01872.

\nolinenumbers


\end{document}